\documentclass[aps,prb,twocolumn,eqsecnum]{revtex4}

\usepackage{graphicx}        % standard LaTeX graphics tool

\begin{document}

\title{Molecular Electronics: From Physics to Computing}
% Use \titlerunning{Short Title} for an abbreviated version of
% your contribution title if the original one is too long
\author{Yongqiang Xue$^{1,*}$ and Mark A. Ratner$^{2}$}
\affiliation{$^{1}$College of Nanoscale Science and Engineering, University at 
Albany-State University of New York, Albany, New York 12203, USA \\
$^{2}$Department of Chemistry and Materials Research Center, 
Northwestern University, Evanston, Illinois 60208, USA \\
{\bf To Appear in Springer Series in Natural Computing} }

% Use the package "url.sty" to avoid
% problems with special characters
% used in your e-mail or web address
%
\maketitle
Even if Moore's Law continues to hold, it will take about 250 years to fill the 
performance gap between present-day computer and the ultimate computer 
determined from the laws of physics alone. Information processing technology 
in the post-CMOS era will likely consist of a heterogeneous set of novel device 
technologies that span a broad range of materials, operational principles, 
data representations, logic systems and architectures. Molecular 
nanostructures promise to occupy a prominent role in any attempt to 
extend charge-based device technology beyond the projected limits of CMOS 
scaling. We discuss the potentials and challenges of molecular electronics 
and identify some fundamental knowledge gaps that need to be addressed 
for a successful introduction of molecule-enabled computing technology.

\section{Introduction}
\label{sec1}
% Always give a unique label
% and use \ref{<label>} for cross-references
% and \cite{<label>} for bibliographic references
% use \sectionmark{}
% to alter or adjust the section heading in the running head
%Your text goes here. Use the \LaTeX\ automatism for your citations

The first functioning transistor was invented by Bardeen, Brattain, and 
Shockley in the late 40s~\cite{BB,Shockley}. It is a bipolar junction 
transistor (BJT) made from a small block of germanium. The first 
integrated circuit (IC) was invented in 1961, which combined monolithic 
bipolar junction transistor and passive components on a single chip. 
This event marked the start of the microelectronics revolution~\cite{IC}. The 
realization of an older transistor principle -- the field-effect transistor (FET) -- 
came about with the metal-oxide-silicon (MOS) transistor in 1962. After the 
development of the complementary MOS (CMOS) circuits, silicon-based 
MOSFETs nearly completely dominated digital logic circuits due to the ease of 
very-large-scale-integration (VLSI) and low power consumption~\cite{MOSFET}. 
For almost 40 years, the VLSI industry has 
followed a steady path of constantly shrinking device geometries and increasing 
chip size, resulting in a history of new technology generation every two to three 
years, commonly refered to as ``Moore's Law''. The 2004 International 
Technology Roadmap for Semiconductors (ITRS) now extends this device 
scaling and increased functionality scenario to the 22-nm technology node 
at year 2016, with projected minimum feature sizes below 10 nanometers 
and chips with more than 6 billion transistors~\cite{ITRS}.  

Equally remarkable with the CMOS scaling is the fact that most of these 
developments have been achieved with the same basic switching 
element (MOSFET), the same basic circuit topology (CMOS), and with a limited 
number of materials (up to about 15 elements in the 1990s). In many 
respects, progress in these areas has been straightford following the design 
scaling rules in the sense that no fundamentally new inventions have been 
needed~\cite{Scaling1,Scaling2}.  However, there 
is no particular reason why Moore's Law should continue to hold: it is a 
law of human ingenuity, not of nature. Indeed, the current ITRS 
roadmap predicts that the scaling of the conventional CMOS technology 
will slow down or stop beyond the 22-nm node. Prior to that time, there 
are a large number of difficult technological challenges at the materials, 
device, circuit and system levels that must be met and overcome, many 
of which currently have no known solutions~\cite{ITRS,Limit1,Limit2,Limit3}. 
This is because nanometersize MOSFET are no longer scaled short-channel 
devices with long-channel behavior. They are true nano-scale devices involving 
the creation and manufacture of objects within the regime of 
nanotechnology~\cite{ERD}. 

In contrast to the CMOS scaling, the technological challenges for 
the information processing industry in the post CMOS era are quite different 
because it is far from clear what needs to be done~\cite{Nano1,Nano2}. 
Nanostructures exhibit a variety of interesting physical, chemical 
and biological properties, many 
of which can be significantly modified by the processing and environmental 
conditions and are not yet fully understood. Progress in innovative device design 
thus often has to come hand in hand with progress in fundamental knowledge
of the physics and chemistry of nanostructures.  Although there is 
a growing consensus that the near term extension of charge-based device 
technology will require a nanodevice technology that is 
architecturally compatible with CMOS and functionally supplementing 
rather than replacing CMOS, questions remain regarding the best direction to 
pursue for such nanoelectronics~\cite{SRC}. 
For example, what are the best functional 
nanostructures, carbon nanotubes, silicon/compound nanowires, or
molecules/polymers? And what are the best device concepts, 
field-effect transistors, single-electron transistors, quantum-effect devices 
(resonant tunneling, quantum interference,...), etc? In the longer term, 
a new nanodevice technology may need to exploit electron and 
electronic charge/current in fundamentally new ways that are closely 
linked to the use of alternate state variables for representing information. 

In general, nanostructured systems may span a broad range of 
materials, data representations, operational principles, and may function 
in different architectures and on different applications. 
Independent of the technology route that post-CMOS devices take,  the 
operation and performance of nanometer-size devices will increasingly 
be governed by atomic-level variations in the materials/device structures 
and processing/environmental conditions. The prospect of device design 
through the bottom-up atom-engineering route of nanotechnology has 
far-reaching impacts on post-CMOS information processing technology either 
as monolithic systems or as polylithic hybrid systems interfacing to the scaled 
CMOS. Molecular nanostructures occupy a prominent role in any 
attempt to offer significant expansion in device functionality beyond the 
end of CMOS scaling. 

In this work, we discuss opportunities for information processing based on 
quantum engineering of the physical states of molecules. Here we define a 
molecule broadly as \emph{a unit whose physical 
(electrical, magnetic, mechanical, optical...) and/or chemical (reactivity, solubility, 
molecular recognition...) properties are sensitive to atomic-scale modification 
of its structure and/or environment}. Note that such a definition of molecule 
essentially covers all nanostructures as defined in the US National 
Nanotechnology Initiative (NNI)~\cite{NNI}, including atoms, organic molecules, 
polymers, nanotubes, nanowires, and nanoparticles, etc. Since the only 
practical general-purpose information processing technology (besides human 
brain) currently available is based on silicon devices which use electron and 
electronic charge/current to drive electronic circuits performing Boolean logic, 
we limit our scope here to charge-based device technology. We don't consider 
device technology options that use fundamentally different physical 
representations of computational state such as optical computing and DNA 
computing. We will not discuss molecular spintronics either which forms 
a separate avenue of research, even though spin is intrinsically associated 
with electrons and nuclei.       

Before we proceed to the main sections of this paper, we want to emphasize 
that our goal here is not to provide a complete survey of molecular electronics 
approaches that have been studied or proposed so far, 
but rather to identify the critial 
research needs in fundamental science that must be addressed in order to 
extend charge-based device technology through the molecular/nano- 
engineering route.  Since any attempt in investigating the potentials of the 
nanotechnology route to information processing should be gauged in reference 
to both its ultimate physical limit and the limits of the ultimately scaled CMOS 
devices, we start with discussions of  the ultimate physical limits to computation 
and the physical factors that account for the success of semiconductor technology 
and their limits in sections II and III respectively.  We discuss the 
main topic of the paper, molecular electronics, in section IV. We conclude 
and summarize in section V.      

%\cite{monograph}.

%\subsection{Subsection Heading}
\section{Ultimate Physical Limits to Computation}

Computers are physical systems, and the laws of physics dictate what they can 
and cannot do~\cite{Landauer,Demon}. Much of the current activity in molecular 
electronics has been motivated by Feynman's pioneering work on 
the physical limits of miniaturization and computation~\cite{Feynman1,Feynman2}. 
Therefore we start the investigation of molecular electronics with examining 
the limits that the laws of physics place on the power of computers. There exists 
a vast literature on this topic~\cite{QC1,QC2}. Our discussion here follows that of 
Lloyd~\cite{Lloyd}, which explores the ultimate physical limits to the computational 
capacity of a computer with a mass of 1 kg occupying a volume of 1 litre (the 
so-called ultimate laptop computer) as determined by the speed of light $c$, 
the quantum scale $\hbar$ and the thermodynamic scale $k_{B}$. 
  
\subsection{Speed Limits}
\label{sec2-1}
A digital computer performs computation by representing information in terms 
of binary digits or bits with logical states $|0 \rangle$ and $|1 \rangle$, and 
then processes that information by performing simple logical operations. 
Any boolean function can be constructed by repeated application of AND, NOT 
and FANOUT, which forms a universal set~\cite{Feynman2}. During such 
logical operations, the bits on which the operation is performed go from 
one state to another. The maximum speed per logic operation can thus 
be determined by how fast a quantum system can move from one 
distinguishable state to another, i.e., the maximum speed of dynamical evolution. 
Since the quantum measure of distinguishable states is the orthogonality of 
states involved, this is best illustrated by considering the minimum time needed 
for the NOT operation, which changes the $|0 \rangle$ state to its orthogonal 
$|1 \rangle$ state or vice cersa. This question is closely related to the 
Aharonov-Bohm interpretation of the time-energy Heisenberg uncertainty principle 
$\Delta E \Delta t \ge \hbar$~\cite{AB1,AB2}: It is not 
that it takes time $\Delta t$ to measure the energy of a quantum system to an 
accuracy of $\Delta E$, but rather that a quantum system with spread in 
energy $\Delta E$ takes time at least $\Delta t=h/4\Delta E$ to evolve to an orthogonal 
state~\cite{Aha}. Instead of expressing the speed of dynamical evolution in terms 
of the standard deviation of energy $\Delta E$, Margolus and Levitin~\cite{ML} 
generalized the result to show that a quantum system with average 
energy E (relevant to its ground state energy) takes time at least 
$\Delta t=h/4E$ to evolve to an orthogonal state.  

Since the simple logical operations of AND, NOT and FANOUT can all be 
enacted in the so-called controlled-controlled-NOT operation~\cite{Toffoli}, 
by embedding the controlled-controlled-NOT gate in a quantum context it is 
easy to show that the maximum speed of logic operation is limited by the energy 
input to the logic gate performing the operation as $4E/h$. More complicated 
logic operations may involve system evolution cycling through a large number 
of quantum states. For evolutions that pass through an exact cycle of N 
mutually orthogonal states at a constant rate, it has been shown that the 
transition time between the orthogonal states is $\Delta t \ge \frac{N-1}{N} h/2E$, 
or the long-sequence asymptotic transition time is twice as 
long as it is for oscillation between $N=2$ states~\cite{ML}.   Applying this result to 
a 1-kg computer with energy $E=mc^{2}$ shows that the ultimate laptop can 
perform a maximum of $4mc^{2}/h \approx 5.426 \times 10^{50}$ 
operations per second~\cite{Lloyd}. 

\subsection{Memory Limits}
\label{sec2-2}
A system with $N$ accessible states can register $log_{2}N$ bits of 
information, so the amount of information that can be registered by a physical 
system is related to its thermodynamoc entropy by $I=S(E,V)/k_{B}ln2$, where 
$S(E,V)$ is the thermodynamic entropy of a system with expectation value for 
energy $E$ confined to a volume $V$.  When it is using all its memory space, 
the ultimate laptop can perform a maximum number of operations per bit per 
second of $\frac{4E}{h}/ \frac{S}{k_{B}ln2} \propto k_{B}T/\hbar $, where 
$T=(\frac{\partial S}{\partial E})^{-1}$ is the operating temperature of the 
ultimate laptop in the maximum entropy state.  A simple estimate of the 
maximum entropy for the 1-kg computer in a litre volume can be obtained by 
modeling the volume occupied by the computer as a collection of modes of 
elementary particles with total average energy $E$, and the maximum entropy 
$S(E,V)$ is that obtained by calculating the canonical ensemble over the 
modes which maximizes $S$ for fixed energy $E$ confined in a fixed volume 
$V$ with no constraint on the spread in energy $\Delta E$~\cite{Beken}. Note 
that this is different from the canonical ensemble normally used for open 
systems that interact with a thermal bath at temperature T. Consequently 
the temperature $T=(\frac{\partial S}{\partial E})^{-1}$ has a different role in the 
context of calculating the maximum entropy of a closed quantum system than 
it does in the case of an ordinary thermodynamic system interacting with a 
thermal bath.

At a particular temperature $T$, the entropy is dominated by the contributions 
from particles with mass less than $k_{B}T/2c^{2}$. The particles contribute 
energy $E=r\pi^{2}V(k_{B}T)^{4}/30\hbar^{3}c^{3}$ and entropy 
$S=2r\pi^{2}V(k_{B}T)^{3}/45\hbar^{3}c^{3}=4E/3T$, where $r$ is the number 
of particles/antiparticles in the species multiplied by the number of polarizations 
multiplied by a degeneracy factor reflecting particle statistics~\cite{Beken}. 
A simple lower bound on the entropy can be obtained by assuming the energy 
and entropy are dominated by black-body radiation of photons, for which case 
$r=2$ (A recent derivation finds the same ultimate limits for information encoded 
using both matter and massless fields~\cite{Lloyd2}). For a 1-kg 
computer confined to 1-litre volume, the maximum entropy state 
corresponds to the operating temperature of $k_{B}T=8.1 \times 10^{-15} J$, 
$T=5.87\times 10^{8} K$. The maximum entropy is $S=2.04 \times 10^{8} J/K$, 
which corresponds to an amount of memory space of 
$I=S/k_{B}ln2=2.13 \times 10^{31}$ bits. When the ultimate laptop is using all 
its memory space, it can perform $4ln(2)k_{B}E/S \approx 10^{19}$ 
operations per bit per second~\cite{Lloyd}.     

\subsection{Thermodynamics of Computation}
\label{sec2-3}
The role of thermodynamics in computation is made clear in the intimate link 
between information and entropy. Ordinary electronic computers are 
thermodynamic engines that do work and generate waste heat. Reducing 
the supply power and removing the heat produced have been main 
technology drivers throughout the history of computing. However, contrary 
to the intuitive thinking, Bennett showed in his pioneering 
paper~\cite{Bennett1} in 1973 that it is possible to construct a general purpose 
computer using only reversible, i.e, one-to-one  logical operations, therefore 
allowing in principle dissipation-less computing if we are willing to compute 
slowly. Energy is dissipated only when information is discarded. Landauer showed 
that irreversible, many-to-one operations such as AND or ERASE require 
heat dissipation of at least $k_{B}Tln2$ for each bit of information 
lost~\cite{Landauer2,Bennett2}. A closely related but separate energy disspiation 
limit has been established for communicating information. Again, in the absence 
of noise, i.e., interaction between the physical system carrying information and 
another uncontrolled physical system, the energy required for transmission of 
a unit of information can be made arbitrarily small if we are willing to do it 
slowly~\cite{Landauer3}. But, as shown by Levitin,  a minimum energy of $kT$ 
must inevitably dissipate in order to transmit a unit of information over a noisy 
channel as a result of the interaction with uncontrolled degrees of freedom 
(environment)~\cite{Levitin}.  More recently, similar fundamental limits on 
the energy transfer associated with a binary switching transition have been 
derived in the context of semiconductor technology by Meindl and Davis~\cite{MD}. 

Besides these fundamental energy dissipation requirements, a realistic computer 
will inevitably be subject to errors during its operation. Error-correcting codes 
can be used to detect these errors and reject them to the emvironment at the 
dissipative cost of at least $k_{B}Tln2$ per bit. Typically such error-correcting 
operations must be done at a high rate in order to maintain reliable 
operation~\cite{Lloyd,Landauer2,Bennett2}. The thermal load of correcting 
large numbers of errors alone can dictate the necessity of operating at a 
slower speed than the maximum allowed by the laws of physics~\cite{Demon,Lloyd}.

\section{Limits of Semiconductor Technology}
The discussion of ultimate physical limits to computation does not 
imply that we can construct a computer that operates at those limits. 
For example, it is inconceivable for present-day technology to control 
computers operating at $T=5.87\times 10^{8} K$, or close to the temperature 
at which electrons and positrons can be produced thermally. Processing, 
storing and transmission of information requires that it be 
represented as the value of some physical quantitity, and physical laws 
control the materials and devices that are used to manipulate 
information~\cite{Keyes1,Keyes2}. Contemporary electronic computers 
operate at speed, memory and energy dissipation capabilities far lower 
than those dictated by the consideration of physical laws alone. From 
the physical perspective, such computers operate in a highly redundant 
fashion. However, there are good technological reasons for such 
redundancy.

\subsection{What Makes a Good Computing Device} 
\label{sec3-1}
Many ingenious proposals for better computing devices were put forward 
and have been the focus of well-funded development efforts as silicon 
microelectronics continued its relentless drive toward miniaturization 
in the past four decades. But the only general-purpose digital computers that have 
ever been built were built with (in the chronological order) electrical relays, 
vacuum tubes, bipolar junction transistors and field-effect transistors. So 
why do so many ingenious schemes fail to realize their promise in electronic 
computation? The answer lies in the vast difference between the conditions 
in which devices are first discovered and demonstrated in the laboratories 
and those in a large system of many devices~\cite{Keyes3,Keyes4}.

For laboratory demonstration of a simple logic circuit, one needs only to choose 
a few proven devices and fine-tune their operating conditions as necessary to 
make them work well to perform a logic operation. But a large computer that 
contains tens of thousands to many millions of devices works in much less 
benign conditions. The output of one device is readily input to another, and 
so on through thousands of step or more.  A large amount of communication 
among the many devices is entailed. There are frequent opportunities 
for a signal to be altered in its passage from one device to another, suffering 
attenuation, diffraction, dispersion and cross-talk on the path. The multiple 
physical and chemical processes used in mass-production of the large 
numbers of component lead to small differences in device characteristics.  
In addition, chemical reactions and diffusion lead to additional unpredictable 
changes in devices over time adding to the uncertainty inherent in manufacturing. 

While the net result of the hazard factors is tolerable in a single logic operation, 
information must pass sequentially through a large number of stages in the 
computing system. Information would soon be lost if the errors introduced were 
allowed to propagate and accumulate from stage to stage. Digital representation 
of information can prevent this by resetting the output of a device to one of the 
standard values after each step. The output of a device may be required as input 
by other devices. The transmission of a signal to a multiplicity of destinations is 
known as fan-out and devices for computers must be able to provide fan-out. 
The standarization of signals and fan-out require that an electrical device 
controls voltage and current larger than those needed to operate it, or a device 
should have both current and voltage gains. Gain is essential to digital devices 
in order that the switching transition at the threshold occupies a small part of 
the signal swing and allows high noise margin. In addition, it is desired that a 
computation in a machine proceed in one direction, from input to final results. 
Each device should operate only on its inputs and not be sensitive to the 
actions or status of the receipients of its outputs. This property is known 
as input-output (I/O) isolation and is required in computer devices. 

The need for I/O isolation, fan-out and high gains put a severe limitation 
on the choice of devices suitable for large computing systems, which was 
only satisfied by the electrical relays, vacuum tubes and transistors. Careful 
examination of other proposed devices showed that they have difficulty 
in satisfying the three conditions simultaneously~\cite{Keyes3,Keyes4}. 
The transistors, especially the silicon MOSFETs, eventually win out due to 
their small size, fast speed, operating stability and low-power consumption.  

The rest of this section is thus devoted to the challenges and limits facing 
semiconductor technology toward the end of the ITRS roadmap as shaped 
by the laws of physics. Such limits can be codified at a hierarchy of levels of 
materials, devices, circuits and systems~\cite{Limit1}.  Many review papers 
have been written on various limits to silicon 
technology~\cite{Limit1,Limit2,Limit3,IEEE1,IEEE2,Max}. We'll focus our discussion 
here only on those aspects of the materials and device limits of silicon 
technology that are likely to be relevant to the CMOS-like route to 
nanoelectronics through molecular/nano- engineering.

\begin{figure}
\centering
\includegraphics[height=5cm,width=8cm]{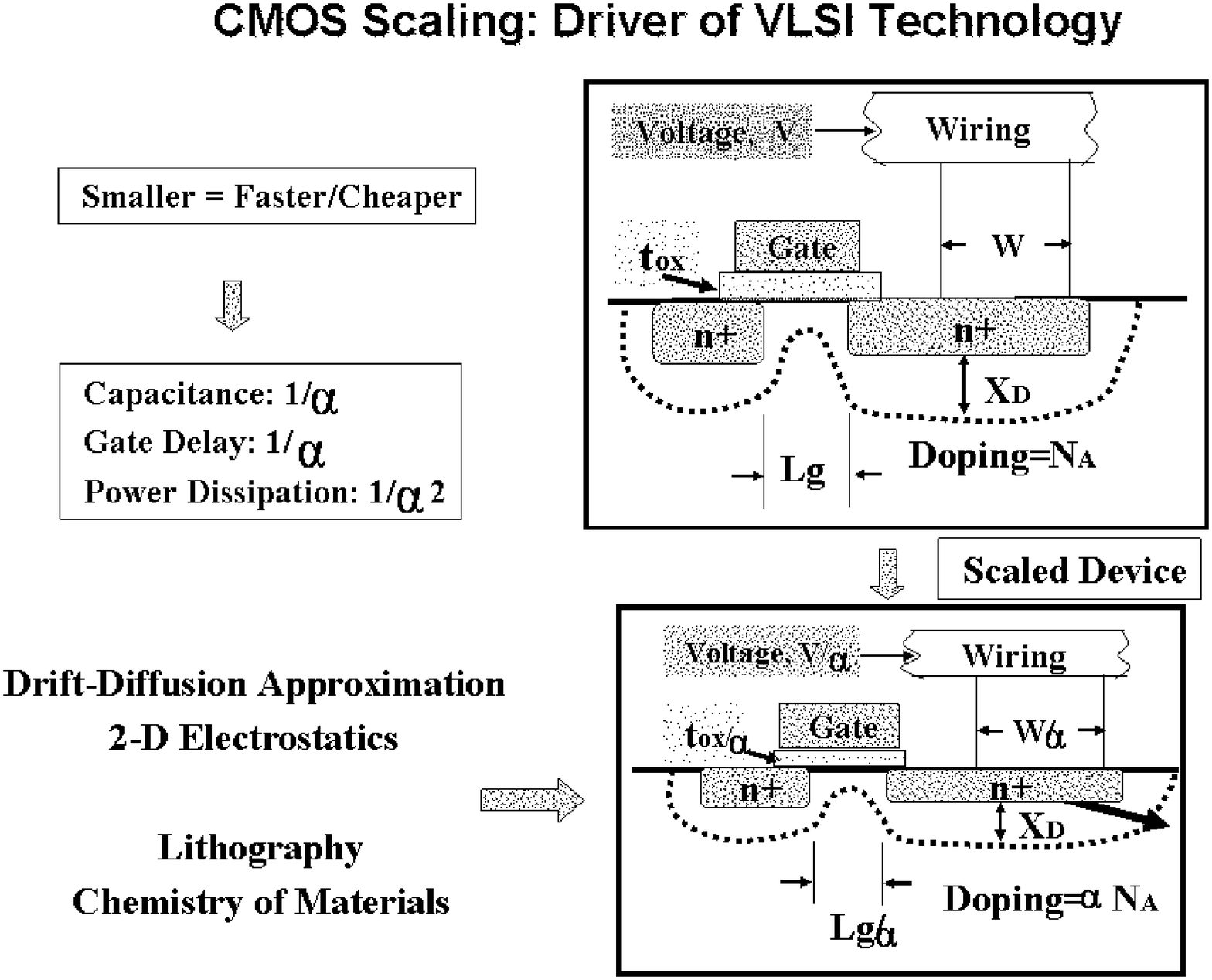}
\caption{Schematic illustration of MOSFET scaling}
\label{xueFig1}   
\end{figure}

\subsection{Materials Limits of MOSFET Technology}
Materials limits are determined by the properties of the particular semiconductor, 
dielectric, and metallic materials used but are essentially independent of the 
structural features and dimensions of particular devices. There are three key 
materials limits: gate stack including both gate dielectric and gate electrode, 
doping in silicon, and contact formation~\cite{IEEE1}. 

The gate insulator in a MOSFET needs to be thin compared to the device channel 
length (a few percent) in order for the gate to exert dominant control over the channel 
potential. But quantum mechanical tunneling of carriers through the insulator 
increases exponentially with decreasing insulator thickness. This puts the 
limit of silicon dioxide ($SiO_{2}$) thickness near 1.0 nm or five atomic 
layers thick for sub-20 nm MOSFET operating at 1V in order to accomodate 
standby power requirments in most IC applications~\cite{Muller}. 
In addition, not all the applied gate voltage is 
efficiently coupled to the channel due to the polysilicon depletion effects in 
the gate electrode and quantum confinement effects in the silicon substrate, which 
add aproximately 0.8 nm to the equivalent electrical thickness in the gate 
capacitor~\cite{IEEE1}.  One solution is to introduce high-$\kappa$ dielectric material 
which reduces the tunneling current while maintaining strong gate electrostatic 
control.  But as the dielectric constant of the insulator increases, 
the band gap tends to 
decrease and the band lineup at the silicon-dielectric interface can be quite 
asymmetric. To avoid thermal emission over a Schottky barrier, a barrier height of 
more than 1V is needed for both electron and hole. Another major barrier any new 
dielectric material will have to overcome is to achieve almost the same low-defect 
density as that of the native $Si-SiO_{2}$ interface. This puts significant constraint on 
the choice of dielectrics and their processing steps. In addition, many of the new 
dielectric materials are unstable in direct contact with silicon and also in the 
presence of the polysilicon gate. Thus it is likely that the entire gate stack will 
have to be replaced with metal gates replacing polysilicon, which has the 
advantage of lifting the gate depletion effect. But polysilicon has the advantage 
that it can be doped either p-type or n-type, shifting the workfunctions so that it 
is suitable for both NMOS and PMOS devices. In contrast, two different gate 
metals are needed for incorporation into a CMOS flow with workfunctions 
near the conduction and valence band edges respectively, which complicates 
enormously the fabrication process~\cite{IEEE1}. 

The second issue is associated with the need for ultra-shallow source/drain 
junctions to reduce the parasitic resistance of the source/drain extension regions 
and the short-channel effect due to drain electric field extending through 
the channel region. This requires increasing the doping density of the source/drain 
region while maintaining abrupt doping profile across the silicon body. However, 
the maximum dopant concentration that can be dissolved in silicon 
under equilibrium conditions (the solid solubility) is $\approx 2\times 10^{21} 
atoms/cm^{3}$) (achievable for arsenic at $\approx 1200 ^{o}C$)~\cite{Doping}. 
Although transient laser annealing can introduce arsenic in metastable electrically 
active concentrations near or above the solubility limit, there is an enormous 
driving force that tends to deactivate the arsenic during any subsequent thermal 
cycling~\cite{Plummer}. The dominant technology used for doping silicon is ion 
implantation, which provides precise control of the placement and quantity of 
doping atoms. But the implantation process produces considerable damage in 
the silicon substrate as a result of the nuclear collisons involved in the stopping 
process. Dopants diffuse by interaction with point defects in the subsequent 
thermal anneal to achieve the desired doping profile. The mechanisms 
underlying the defect formation and dopant diffusion process are far from 
being fully understood~\cite{DefectRMP}. 

The third issue is associated with the junction contact formation. Contacts 
in silicon technology are normally made with self-aligned silicides containing 
heavily doped silicon. This process provides an ohmic contact covering the 
area of the source/drain diffusion and minimizes the contact resistance. Further 
reducing the contact resistance with decrease of feature size requires increasing 
the silicon doping and reducing the Schottky barrier height. The doping is limited 
by the solid solubility as discussed earlier. Barrier height engineering in 
metal-silicon system remains not fully understood despite its obvious technical 
importance. In addition, the silicide formation process consumes the top portion of 
silicon as the metal is deposited and reacted to form the silicide, this can 
increase sheet resistance of the source/drain extension region and also 
change the dopant structure adjacent to the metal. 
 
\subsection{Device Limit of MOSFET Technology} 
Historically MOSFET scaling has been governed by the need to preserve the 
good electrostatic behavior at the reduced device dimension, i.e., reducing 
supply voltage and gate insulator thickness and increasing doping 
concentration.  The traditional limit of device scaling is determined thus by 
the effects that modify the ideal electrostatic contol. These include quantum 
effects due to tunneling leakage through gate insulator, tunneling through 
body-to-drain junction,  
direct source-to-drain tunneling, thermal effects due to thermally generated 
subthreshold current at room temperature and also the increasing sensitivity to 
minute fabrication spreads. In addition to such limits intrinsic 
to small device size, other limits more intimately connected to the materials 
and device structure of the ultimately scaled MOSFET have been proposed 
which, as accutely pointed out by Fischetti, suggests changing the ``nature'' of the 
nanometer-size MOSFETs moving toward the sub-10nm regime~\cite{Max}. 

The most fundamental one seems to be that set by the long-range Coulomb 
interaction between the channel electron and the ``high-density'' electron gas 
in the highly doped source, drain and gate electrodes. This is reflected in: (1) 
the emission and absorption of the low-frequency plasmon (on the order 
of magnitude of $meV$) in 
the source/drain by the channel electrons which thermalizes the hot-electron 
distribution in the channel and indirectly reduces the effective electron velocity; 
(2) the ``Coulomb'' drag between the channel electron and electrons in the gate 
(also plasmon-mediated) across the very thin insulator results in a direct 
loss of momentum of the channel electrons. Both effects may contribute to 
the breakdown of ``ballistic'' transport widely assumed in current theoretical 
estimates of the MOSFET scaling limit~\cite{Ball}: Short channel is required for 
``ballistic'' transport, but the increased strength of Coulomb interaction may  
kill it at the outset~\cite{Max}. Combining with other less fundamental but 
equally important effects such as ``remote'' phonon scattering in the gate 
stack and scattering accompanied with substrate engineering, this may 
contribute further to the end-of-the-road scaling scenario that there may 
not be a single end point for scaling, but instead many end points, each 
adapting optimally to its particular applications~\cite{IEEE2}.              

\section{Molecular Electronics: From Physics to Computing} 

\subsection{Motivation and Definition}

\emph{Even if Moore's Law continues to hold}, it will take about 250 years of 
exponential scaling to fill the gap between the ultimate laptop operating 
on $10^{31}$ bits at $10^{51}$ operations per second and the present-day 
laptop operating on $10^{11}$ bits at $10^{10}$ operations per second. 
Although the ultimate laptop operates at conditions that do not seem 
to be contollable at all from present-day technology, new physical principles may 
be imagined that turn today's inconceivable into tomorrow's common sense 
if we remember quantum physics has only 100 years' history. But we are not 
concerned with such exotic possibilities beyond the horizon of current 
understanding of physical laws. The technological goal of molecular electronics is 
instead to extend the performance increase of charge-based 
device technology beyond that perceivable from CMOS scaling at the 
projected end of ITRS roadmap as far as possible, based on innovative utilization 
of functional nanostructures and quantum mechanical laws. 

Although many technological barriers exist for which there are currently 
no known solutions, the past success of CMOS scaling gives us all reason to 
believe that the projected goal of the CMOS scaling at 2016 will be 
surpassed~\cite{IEEE3}, at which point the ultimate MOSFET will have gate oxide 
thickness in the 1.0-nm range, channel thickness in the 3.0-nm range, and 
channel length in the 9.0-nm range~\cite{Note}. Since CMOS technology is 
the only practical general-purpose information processing technology 
(besides human brain) currently available, investigation along the molecular 
electronics route should be gauged in close reference to the continually 
scaled CMOS devices in both its conventional and ``nonclassical'' 
forms~\cite{CMOS1,CMOS2}. In addition, the future devices and their 
target performance metrics should meet the generic criteria of: (1) that they 
need to be of high performance in terms of speed and density while remaining 
energy efficient; (2) that they should be structurally stable under room 
temperature operation and not be dominated by parametric variations due to 
processing and environmental conditions; (3) that they should be scalable through 
multiple generations with integer multiples of performance. In the near term, 
they might preferably be capable of integration on a CMOS platform, but the long 
term options should be kept open (remember the 250-year span!). Consequently 
we shall consider materials and device issues associated with 
both molecular/nano- engineered devices that are structurally and/or functionally 
similar to CMOS devices (referred to as CMOS route hereafter) and 
molecular/nano- engineered devices that are configured for information 
acquisition, sensing, storage and transmission in ways fundamentally different 
from the CMOS devices (referred to as Non-CMOS route hereafter).

In the preface to the first edition of his widely popular textbook on 
semiconductor devices published in 1969, Sze defined a semiconductor 
device as \emph{a unit which consists, partially or wholly, of semiconducting 
materials and can perform useful functions in electronic apparatus and 
solid-state research}~\cite{Sze}. Correspondingly we define a molecular 
electronic device as \emph{a system which consists, partially or wholly, 
of individual molecules and can perform useful functions in electronic 
apparatus and nanostructure research through atomic-scale control}. 
We only discuss molecular electronics for applications in information 
processing device here and leave the discussion of molecular electronics 
as ``artificial'' laboratory of nanoscopic physics 
for other efforts~\cite{Xue1,Ratner}.  

\subsection{Molecular Electronics: CMOS Routes} 

{\bf Molecular Transistor}

Three-terminal devices, i.e., transistors, have been indispensable for building 
digital logic systems based on semiconductor technology due to the stringent 
requirement of I/O isolation, large noise margin and signal gain. Molecular 
field-effect transistors (MolFET), where the active part of the device is 
composed of quasi-one-dimensional (Q-1D) nanostructures like carbon nanotubes 
or nanowires, have been widely studied that are structurally and functionally 
similar to their CMOS analog~\cite{CNTFET1,CNTFET2,CNTFET3,NWFET1,NWFET2}.  
Q-1D nanostructures offer additional advantage as alternative channel 
materials in the CMOS route since they can function both as active devices 
and interconnects and thus have the potential to provide simultaneously 
two of the most critical functions in any integrated 
nanoelectronics~\cite{CNT,NW1,NW2}. Experimental progress on single devices 
has been fast, and useful simple circuits like inverter, mixer and decoder have been 
demonstrated~\cite{CNTC1,CNTC2,NWC1,NWC2,NWC3,NWC4}. There are 
many points of confluence between the technologies of the scaled 
silicon devices and Q-1D nanostructured junctions and transistors, including the 
integration of high-$\kappa$ gate stack, homo(pn)- and hetero- junction 
diodes and transistors, substrate engineering (strain) and ``non-classical'' 
transistor structures~\cite{NW1,NW2,CNTD1,CNTD2}. Investigation along this route 
provides thus an ideal reference point both for exploring novel device design 
at the molecular scale and for re-examining the physical principles of semiconductor 
microelectronics from the bottom-up approach. Here carbon nanotube and 
semiconductor nanowire offer subtle but significant differences in their prospect 
for post-CMOS information processing. 

\begin{figure}
\centering
\includegraphics[height=5.0cm,width=8cm]{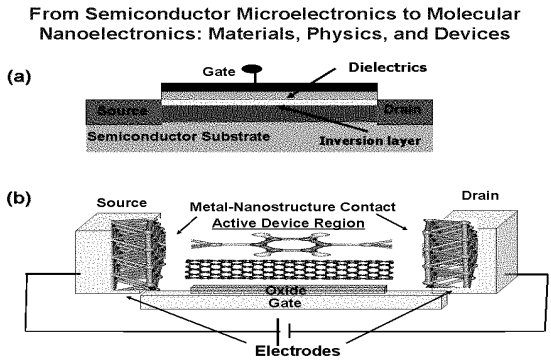}
\caption{Schematic illustration of the device structures of a conventional 
CMOS device and a typical nanodevice.}
\label{xueFig2}   
\end{figure}

Carbon nanotubes (CNT) are hollow cylinders composed of one or more concentric layers 
of carbon atoms in a honeycomb lattice arrangement, which typically have 
a diameter of 1-10 nm and a length of several nanometers to several micrometers. 
In addition to the small size, CNTs offer some salient features that make them 
attractive  candidates for electronic devices~\cite{CNTFET2}: (1) 
The quasi-1D structure implies a 
reduced phase space for carrier scattering by both impurity and lattice 
vibration. It also leads to distinctly different electrostatic behavior from the planar 
silicon device which affects both screening and tunneling. (2) The C-C 
$sp^{2}$ bonding leaves no dangling bond on the surface. In particular, for 
single-wall carbon nanotube (SWNT) all carbon atoms are surface atoms. CNT 
electronics are not bound to use $SiO_{2}$ as an insulator and novel transistor 
structures like surrounding gate transistors can be adapted. (3) The strong  C-C 
$sp^{2}$ bonding gives CNTs high mechanical and thermal stability. Current 
densities $\ge 10^{9} A/cm^{2}$ can be sustained. Several critical issues 
related to contact, doping and scattering remain to be sorted out for further 
development of CNT-based nanoelectronics. 

In contrast to silicon MOSFET, the source, drain and gate 
electrodes in MolFET are currently made from deposited or lithographically defined metals. 
The Schottky barriers at the CNT-metal contacts play a significant role in 
determining the transport characteristics~\cite{CNTFET2,CNTB1,CNTB2} (we can also 
expect that the Schottky barrier problem will play an increasingly important role 
as MOSFET scales toward sub-10nm regime, since the low-frequency plasmon 
in the doped source/drain region can be removed by using metal electrodes). 
Due to the Q-1D geometry, both the barrier height and barrier shape are 
important in determining the relative 
importance of tunneling and thermionic emission across the barrier. The
recent observation of ohmic contact using Pd provides a particular 
challenge~\cite{CNTD1,CNTD2} as the previous theoretical study shows 
similar Schottky barrier for Pd and Au that have similar work functions. 
However, the model used assumes only electronic coupling across the 
interface with fixed atomic structure. Transition metals including both Ti 
and Pd are known to be chemically active attaching to CNT surface and 
can form carbide immediately adjacent to the interface.~\cite{CNTFET1,CNTFET2} 
Recent experiments have also shown that Schottky barriers can be significantly 
lowered by chemical treatment of the metal-CNT interface~\cite{CNTCh}. 
Work will be needed to extend the theoretical model for better study of the 
interface chemistry including structural relaxation effects in the configuration 
of CNTFET with different gate structures. 

Doping in a semiconductor typically implies introducing a shallow impurity atom 
into the host lattice using ion implantation or thermal diffusion accompanied by 
creation of lattice defects~\cite{DefectRMP}. But it may take a fundamentally 
different approach in CNTs. For example, doping in carbon nanotubes can be 
introduced chemically by exposing the CNT surface to alkali metals, 
by inserting $C_{60}$ molecules inside the CNT, by surface functionalization 
with molecules/polymers for charge-transfer doping (which is essentially 
the electronic basis of sensing). In addition, the doping type can be converted 
between p-type and n-type by chemical treatment using e.g. oxygen and 
molecular hydrogen~\cite{CNTFET1,CNTFET2}. Doping in nanotubes can also 
be introduced physically using electrostatic gating or contact-induced charge 
transfer~\cite{CNTPN}. ``Self-doping'' mechanisms for intrinsic SWNT caused by 
curvature induced charge redistribution have also been proposed, which shift 
the Fermi-level position inside the band gap~\cite{CNTSD}. Despite its 
obvious importance, comprehensive experimental and theoretical study 
and a coherent  physical picture of the various doping mechanisms, including 
both electronic and structural consequences, have not yet appeared.  
A particularly interesting question in this regard is the optimal doping limit 
in carbon nanotubes for both physical and chemical doping mechanisms.

The major scattering mechanisms in CNTFET are those due to defects including 
dopant, gate stack and phonon. Due to the reduced phase space, the 
probability of back-scattering by defects and acoustic phonon is significantly 
reduced at low-bias compared to the planar silicon 
devices~\cite{CNTPh1,CNTPh2,CNTPh3}. The absence of 
reactive dangling bond states at the CNT surface also make it less likely to 
suffer significant scattering due to the interface states and charge traps at the 
channel-gate interface. But it remains unclear how these favorable conditions 
may be modified at high-bias. These include optical phonon emission by the 
energetic carrier, the injection of carriers into the gate dielectric and the resulting 
gate insulator degradation, remote phonon scattering between channel electrons 
and gate phonons, and the structural stability adjacent to the intrinsic or doping 
induced defect site. The optical phonon scattering length has been estimated 
at $\approx 10$ nm~\cite{CNTPh2}, but in the absence of a realistic quantum 
transport model of electron-phonon coupling in CNTFET, this result should be 
taken with reservation~\cite{CNTPh4}. 
Many fundamental knowledge gaps  need to be addressed before we can 
have a convincing picture of the performance limit of CNTFETs in comparison to 
that of the ultimately scaled MOSFET. The recent report on suspended carbon 
nanotubes  seems to suggest a cleaner platform for investigating many of the 
issues involved~\cite{CNTNew}. 

A different scenario applies to the nanowire FETs (NWFET), which seem to be 
less controversial. The vapor-liqiuid-solid phase growth process using 
nanoclustered catalyst pioneered by the Lieber group has led to the 
fabrication of single-crystal silicon nanowires~\cite{NW1}, 
where the size distribition of 
the nanowires is determined by that of the catalyst nanoclusters.  Both n-type 
and p-type dopants can be selectively inserted during the nanowire growing 
process. This has opened up the scheme of fabricating complementary logic 
circuits on the single silicon nanowire, where source/drain electrodes can be 
lithographically defined after the doped segments have been grown. Since the 
diameter of the nanowires is typically several tens of nanometer, well-known 
techniques in forming metallic contact in planar silicon device can be adapted 
leading to low barriers and low resistance contacts~\cite{NW1,NW2}. More recently, 
innovative techniques have been reported that solve the integrated contact 
and interconnect problem through selective transformation of silicon nanowires 
into metallic silicide nanowires~\cite{Lie1}. The single-crystal metallic silicides 
have high conductivity and high failure current, while being capable 
of forming atomically sharp metal-semiconductor heterostructures with the 
silicon nanowire of 
similar diameters. This opened up the possibility of an ultra-dense integrated 
nanosystem that integrates both active device area and high-performance 
interconnect from a single nanowire building block while benefitting from the 
knowledge gained in the planar silicon devices (in particular the 
silicon-on-insulator approach) with minor modifactions. In 
addition, different elemental, binary and ternary nanowires can be fabricated 
using the same vapor-liquid-solid geowing process, providing a significant 
design freedom for system designers~\cite{NW1,NW2}. 

Both carbon nanotube and nanowire field-effect transistors have been demonstrated 
showing favorbable performance compared with the state-of-the-art 
silicon MOSFET, while leaving substantial room for materials and device design 
optimization. Carbon nanotubes, even though of much smaller diameter than 
silicon nanowires, do not have the advantage of integrated metallic ocntact 
on the single-tube basis. This is because the reduced phase space and 
the correspondingly low electron density of states in the metallic SWNT 
do not allow rapid relaxation of carriers injected through the channel, 
which has to be connected to a larger area metal electrode to allow 
I/O separation and efficient heat removal.   Athough this may be remedied 
by using bundles of metallic SWNT or metal nanowires, further materials 
and fabrication challenges need to be resolved in addition to the Schottky 
barrier problem in such interfaces.  The challenge  for nanowire FETs is 
instead to scale the nanowire to true molecular dimension while maintaining 
scalable performance gain~\cite{Lie2}.

{\bf Molecular Interface to CMOS}

Direct integration of molecular functionality with the scaled CMOS technology 
forms a starting point for hybrid top-down and bottom-up approaches. Such hybrid 
approaches may combine a level of advanced CMOS lithographical design pattern 
that represents designer-defined information and a level of molecular structures 
self-assembled with great precision and functional flexibility. This combines 
the advantages of nanoscale components, such as the reliability of CMOS 
circuits and the minuscle footprints of molecular devices, and the advantages 
of patterning techniques, such as the flexibility of traditional photolithography 
and the potential low cost of nanoimprinting and chemically directed self-assembly, 
to enable ultra-dense circuits with acceptable fabrication costs. 

One promising direction is to use molecules as charge storage elements for 
nonvolatile memory in the MOSFET structure. Nanocrystal and quantum-dot 
memories are examples of flash memories that utilize quantum dots between the 
gate and the channel of the field effect transistor to store electrons, which 
screen the mobile charge in the channel, thus inducing a change in the 
threshold-voltage or conductivity of the underlying channel~\cite{Dot1,Dot2,Dot3} 
The quantum dots are isolated 
from the gate, and their processing can be accomplished together with 
CMOS processing. Both metallic and semiconductor nanocrsytals embedded 
in the gate oxides have been explored, but to enable reliable operation utilizing 
the single-electron effect at room temperature, truly molecular 
dimension ($\approx 1 nm$) quantum dots are preferred.    
    
Recent work has demonstrated the integration of fullerenes including $C_{60}$ 
and $C_{70}$ in the gate stack of CMOS technology~\cite{MolDot1,MolDot2}. An 
electrically erasable programmed read-only-memory (EEPROM) type device 
was fabricated by effecting molecular redox operations through non-volatile 
charge injection, which occurs at a specific potential of the 
fullerene molecules with respect to the 
conduction band of Si at the $Si/SiO_{2}$ interface. Compared to metal and 
semiconductor nanocrystals which have non-negligible size variations, the 
mono-disperse nature and small size of fullerene molecules lead to large and 
accurate step-wise charging into the molecular orbitals and may potentially 
provide reliable muti-level storage with electrostatic control. 

Alternatively, the body thickness control in the quantum-dot memory can be 
solved using CNTFETs which have monodisperse nanoscale cross sections. 
A new nonvolatile memory structure has been reported which uses a back-gated 
CNTFET as sensing channel and metal nanocrystals embedded in the dielectric layer 
near the SWNT as charge storage media~\cite{MolDot3}. The gate electrode 
regulates the charging and discharging of the metal nanocrystal, which imposes 
a local potential change on the nanotube channel and alters its electrical 
conduction. The device shows clear single-electron sensitivity and Coulomb 
blockade charging~\cite{MolDot3}. 

A closely related concept is to use redox-active molecules self-assembled on 
nanowire field-effect transistors for nonvolatile memory and programmable 
logic applications~\cite{MolDot4}. Multi-level molecular memory 
devices have been demonstrated using porphyrin molecules self-assembled 
on $In_{2}O_{3}$ nanowire transistors for nonvolatile data storage up to 
three bits per cell~\cite{MolDot5,MolDot6}. Charges were placed on the redox 
active molecule. Gate voltage pulses and current sensing were 
used for writing and reading operations. Here replacing the gate insulator layer with 
self-assembled molecular components reduces significantly the device size, which 
simplifies fabrication and may aviod potential damage to the molecular component 
during the gate stack formation. In addition, different molecule-nanowire combination 
may be chosen leaving enormous room for design optimization. This seems to be a 
very promising direction, although many fundamental questions regarding the nature 
of the molecular states during read and write operation remain to be sorted out.  

\subsection{Molecular Electronics: Non-CMOS Routes} 

{\bf Molecular Switch}
 
The situation for designing three-terminal switching devices on the molecular scale 
becomes much less clear once we move out of the proven domain of CMOS-like 
information processing~\cite{Joachim}. This is exemplified by the lack of field-effect 
transistor effect in devices made from short ( $\approx 1 nm$) molecules, since effective 
gate control requires the placement of gate in close proximity to the molecule 
(a few angstrom away) while avoiding overlap with the 
source/drain electrodes~\cite{MolFET1}. One approach to demonstrate  
strong gate control in such small scale is to use an electrochemical gate 
by inserting the device in electrolytes. Here the gate voltage falls mostly across 
the electrical double layer at the electrode-electrolyte interface which is only 
a few ions thick, and strong field effects on the source/drain curent have been 
observed for a perylene tetracarboxylic diimide molecule $2.3$ nm long covalently 
bonded to two gold electrodes at gate voltage of $-0.65 V$ due to the field-induced 
shift of molecular orbitals relative to the electrode 
Fermi level~\cite{MolFET2}. 
However, further increasing gate voltage causes the device to break down. The 
electrochemical gating techniques has also been applied to CNTFETs~\cite{MolFET3}, 
but the scaling characteristics of such electrochemical transistors remains unknown.  

Another way of achieving a strong field regulation effect is to put charged species 
in close proximity to the molecules. One recent experiment demonstrated the 
modification of current-voltage characteristics through a single-molecule in a 
STM junction by nanometer-sized charge transfer complex, where the electron 
acceptor is covalently bonded to the junction molecule and the electron donor 
comes from the ambient fluid. The effect was attibuted to an interface dipole 
which shifts the Fermi level of the substrate relative to the molecular 
orbitals~\cite{MolFET4}.  Another approach used scanning tunneling microscope 
(STM) contact to styrene-derived 
molecules grown on a $Si(100)$ surface. The strong field effect arises from 
charged dangling bond states on the silicon surface, the electrostatic field of 
which shifts the molecular levels relative to the contact Fermi level. 
The effect can be modulated by STM manipulation of the surface charging 
state or the molecule-charged centre distance~\cite{MolFET5}.  
   
Switching by mechanical movement of an atom in the molecule has been 
proposed for a long time.  An ingenious purely mechanical computer has recently 
been demonstrated by researchers from IBM, which was made by creating a precise 
pattern of carbon monoxide molecules on a copper surface~\cite{Cascade}. 
Tiny structures, termed ``molecular cascade'', have been designed and 
assembled by moving one molecule at a time using an ultra-high-vacuum 
low-temperature STM, that demonstrated 
fundamental digital logic OR and AND functions, data storage and retrieval, 
and the``wiring" necessary to connect them into functioning computing 
circuitry. The molecule cascade works because carbon monoxide molecules 
can be arranged on a copper surface in an energetically metastable 
configuration that can be triggered to cascade into a lower energy 
configuration, just as with toppling dominoes. The metastability is due to 
the weak repulsion between carbon monoxide molecules placed only one 
lattice spacing apart. 

To overcome the intrinsically slow speed due to atomic/molecular motion, 
a molecular 
electro-mechanical switch has been proposed. An early suggestion of atomic relay 
transistor proposed to use the mechanical motion of an atom to cause 
conductance change or switching of an atomic wire~\cite{ART1}. Theoretical 
calculations suggest high switching speed of $\ge 30$ THz or $\ge 100$ Thz 
if a silicon or 
carbon atom is used as the switching atom respectively, where a displacement 
of the switching atom by only one diameter would change the conductance 
of the atomic wire by orders of magnitude~\cite{ART2,ART3}. Such an atomic 
relay transistor was recently demonstrated using electrochemical gate control 
of silver atoms within an atomic-scale junction~\cite{ART4}. A switching time of less than 
$14 \mu S$ was estimated. An early molecular version of electro-mechanical 
amplifer was demonstrated using STM manipulation of $C_{60}$ molecules, 
where current flowing through the $C_{60}$ molecule can be modified 
exponentially upon minute compression of the molecule by the STM 
tip~\cite{ME1}. More 
recently, a molecular version of the atom relay transistor has been demonstrated 
based on the rotation of the di-butyl-phenyl leg in a 
Cu-tetra-3,5 di-tertiary-butyl-phenyl porphyrin molecule, where 
the intramolecular motion of the switched leg is controlled mechanically by the 
tip apex of a noncontact atomic force microscope~\cite{ME2,ME3}. 
The comparison of the experimental and computed forces shows that rotation 
of the switched leg requires an energy of less than $100\times 10^{-21} J$, 
or four orders of magnitude lower than the state-of-the-art MOSFET.    

The three-terminal switching devices just discussed, although ingenious 
and scientifically provoking, do not seem to satisfy the requirements of I/O 
separation, gain and fan out for digital applications and there is no known scheme 
for extending them to large scale integration. Several two-terminal molecular 
switching devices have been proposed and demonstrated based 
on the reversible conformational change upon application of an 
electrical field~\cite{Switch1, Switch2,Switch3,Switch4,Switch5,Switch6}. Different 
mechanisms have been proposed for such bistable molecular 
devices~\cite{Switch4, Switch5,Switch6,Switch7,Switch8,Switch9,Switch10,Switch11}. 
Other bistable devices showing negative differential resistance have also been 
observed~\cite{NDR1,NDR2}. The two-terminal bistable devices have a long 
history in solid state electronics including in particular tunneling and 
resonant tunneling diodes based on semiconductor homo- and hetero- 
junctions~\cite{Sze2}. Despite the enormous efforts 
put into logic design using two-terminal devices, sucess is limited~\cite{RTD}. And 
it is now well known that the bistable characteristics is unfavorable for large 
computing system in many ways~\cite{Keyes3,Keyes4}. The critical point is that 
gain in the bistable logic depends on biasing the circuit  close to the threshold 
so that the addition of only a small input can cause a large change in the output. 
This puts great demand on the precision with which this can be done and gain 
is hard to realize in a noisy world with variable components. In addition, there is 
no standardization of signal values and there is no convenient inversion operation. 
This has forced research innovations in molecular electronics 
architecture~\cite{Design1,Design2}.   
Similar objections apply to cellular automata type devices, for which molecules 
have been suggested for optimal implementation~\cite{Keyes3,Keyes4}. In the 
cellular automata approach, connecting devices together by wiring is avoided 
by letting each device interact directly with its nearest neighbours. Previous 
research suggests that the capabilities of cellular automata in large computing 
systems are limited: they do not allow efficient execution of frequent access to 
memory and branching to other computational routines because interactions 
with distant information occur by shifting data one step at a time. It is not clear yet 
how much advantage molecular self-assembly can bring to cellular automata 
or other collective computing paradigms~\cite{Collect}. 

{\bf Molecular Single-Electron Devices} 
 
Single-electron devices - in which the addition or subtraction of a small number 
of electrons to very small conducting particles can be controlled at the 
single-electron level through the charging effect - have attracted much 
attention from the semiconductor industry as an alternative device technology 
that could replace CMOS beyond the 10-nm frontier~\cite{SETBook,KK1,KK2}. The 
previous discussion of molecular quantum dot memory has highlighted the 
potential advantage of molecular components in single-electron memories. 
For logic applications, molecular implementation of single-electron transistors is 
equally important since molecular-scale field effect transistors cannot help solve the 
key problem of transistor parameter sensitivity to channel length. Research in the 
past decade shows that there are two major obstacles preventing the wide-spread 
application of single-electron logic: (1) the need to operate at very low 
temperature; and (2) the ultra sensitivity to background charge noise. 

The potential size advantage of molecular components to enable room-termperature 
operation is obvious. Both theory and experiment show that for reliable 
operation of most digital single-electron devices, the single-electron addition
energy ($E_{C}$) should be approximately 100 times larger than $kT$~\cite{KK2}. 
This means that for room temperature operation, $E_{C}$ should be as large as 3 eV, or 
quantum-dot size about 1 nm.  Molecular electronics offers a solution to this scaling 
limit by taking advantage of the bottom-up self-assembling process.  In addition, 
using molecules with precise chemical composition may potentially solve the 
reproducibility problem in conventional metal/semiconductor clusters or 
electrostatically defined quantum dots in two-dimensional-electron-gas (2DEG) 
due to the size and shape fluctuations. Note that single-electron effects 
have also been demonstrated using carbon nanotubes, but their larger size 
makes them less likely candidate for reliable room temperature 
operation~\cite{CNTSET1,CNTSET2}. The solution of the random background 
charge problem is much more difficult. Note that the electrostatic potential 
associated with random charged impurities in the environment is a problem 
for any nanoscale devices. But it poses a particularly tough problem for 
single-electron devices because of their large charge sensitivity. 
 
A comparison between the conventional approach and several representative 
single molecule-based single-electron devices shows clearly the new 
physical processes introduced by the use of molecular-scale 
components~\cite{MolSET1,MolSET2,MolSET3,MolSET4,MolSET5,MolSET6}.  
The molecular-scale dimension of the quantum dot leads to two intrinsic 
effects due to the ultra-small size: (1) both the wave function and the 
energy of the discrete electron states of the quantum dot depend on 
the size, shape and net charging state of the quantum dot; (2)  Due to 
finite number  of degrees of freedom and lack of an efficient relaxation 
mechanism on the quantum dot, the quantum dot may stay in non-equilibrium 
state and self-heating may occur during the cycle of single-electron transfer. 
In addition, as electrons are added or removed from the molecular quantum 
dot, both the shape of the molecule and its position relative to the contacts 
may be altered. The electron state of the molecular-scale component is also 
sensitive to the atomic-scale change of the environment, e.g. due to presence 
of surface states which in turn may be modified by surface adsorption, the 
presence of impurities on the contact surface and/or the interaction with 
neighbor quantum dots. Treatment of all the above processes goes beyond 
the conventional theory of single-electron tunneling and is important for 
quantitative and realistic evaluation of their figures of merit. 

\begin{figure}
\centering
\includegraphics[height=5.5cm,width=8cm]{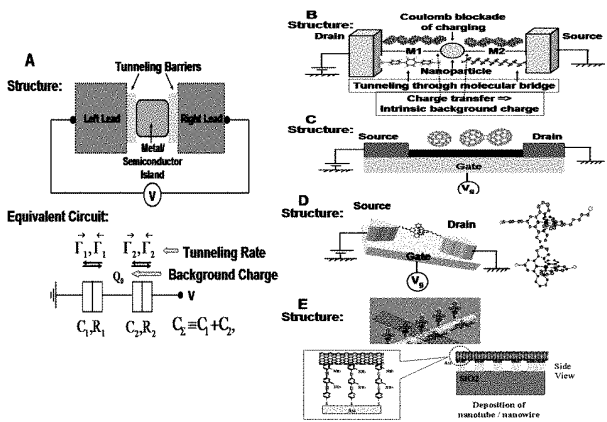}
\caption{A, Typical structure and equivalent circuit of the conventional 
single-electron devices. B, Self-assembled or bio-directed assembly of 
single-electron device fabricated through synthetic routes. The nanoparticles 
are connected to the electrodes and/or to each other through either organic 
linkers or biomolecules with molecular-recognition capability~\cite{MolSET1,Xue}. 
C, A quantum dot is formed by a single C-60 or C-140 molecule physisorbed between two 
metal electrodes~\cite{MolSET2,MolSET6}. The molecule may start oscillating 
as discrete charges are added to or extracted from the molecules through the 
contact. D,  The quantum dot is a single metal atom embedded within 
a larger molecule and connected to the metal contact pads through 
insulating tethers~\cite{MolSET3,MolSET4}. E, The 
molecule can also be adsorbed on top of a nanowire transistor which 
provides the source/sink of single electrons~\cite{MolDot5}.    }
\label{xueFig3}   
\end{figure}      

So far, these devices have been formed by techniques excluding practical 
fabrication of integrated circuits. But there are good prospects for
chemical synthesis of special molecules that would combine the structure 
suitable for single-electron tunneling with the ability to self-assemble from
solution on prefabricated nanostructures with acceptable yield, opening a 
way to generically inexpensive fabrication of VLSI circuits. For logic circuits, 
the random background charge effects remain hard to overcome. 
Nevertheless, it has been suggested that the hybrid molecule-CMOS circuits, 
or ``CMOL'' circuits, that combine CMOS stack with molecular single-elctron 
devices interconnected by nanowires, in defect-tolerant architectures that 
allow to either tolerate or exclude bad devices, may become 
the basis for implementation of novel, massively parallel architectures for 
advanced information processing, e. g., self-evolving neuromorphic 
networks~\cite{KK2}. Such hybrid approach can help to solve the low gain of 
single-electron transistors, but it remains to demonstrate reliable 
high-performance digital circuits. 

{\bf Molecular Quantum-Effect Devices}

Intensive research on semiconductor heterostructures in the past three decades 
has generated many novel device concepts based on tunneling, resonant 
tunneling, real-space transfer, hot-electron transport and quantum wave 
interference effects, in addition to creating the entire field of mesoscopic 
physics~\cite{Hetero1,Hetero2,Hetero3,Hetero4,Hetero5}. Although they have yet not  
generated a real breakthrough in microelectronics, quoting a sarcastic statement 
from the mainstream silicon community, ``heterostructure is and will be the 
material of the future'', they provide a foundation and rich source of inspiration 
for going beyond the limits of conventional devices through quantum engineering 
of physical states in confined systems~\cite{Junc1,Junc2,Junc3,Junc4,Junc5}. 
Recently they are also subjects of rejuvenated 
interest as MOSFET moves toward the sub-10 nm era based on advanced 
silicon-on-insulator (SOI) structures and $Si-SiGe$ heterostructures~\cite{Luryi}.  

Molecules are intrinsically heterostructures. Molecular electronics offer the 
ultimate testing ground for quantum-effect devices based on the atom-engineering 
approach to the heterostructure concept. Research in this field is intimately 
connected to exploiting molecular electronics as artificial laboratory of new 
principles of nanoscopic physics~\cite{Xue1, Xue2}. This is still a vaguely defined 
area and much fundamental knowledge needs to be sorted out. But molecular 
heterostructures already offer multiple device opportunities that are beyond 
the capability of or at least very difficult to achieve in scaled silicon devices. 
In the case of Q-1D nanostructures, this includes the possibility of fabricating 
metal-semiconductor and semiconductor heterojunctions with simultaneous 
band-gap engineering on a single nanotube and nanowire basis, and the possibility 
of fabricating Y-junction, T-junction, branched nanowires and superlattice 
devices with atomically sharp interfaces~\cite{NW1,
NW2,Lie1,Lie2,NW2,Xu,NWNew, CNTN1,CNTN2,CNTN3,CNTN4,CNTN5,CNTN6,CNTN7}. 
Similar quantum-effect devices can also be implemented on single-molecule 
basis through a synthetic chemistry approach, but can involve very different 
physical mechanisms and operation principles~\cite{Joachim,MolFET4}. Some examples 
are single-molecule heterostructures where saturated molecular groups can 
be selectively inserted between molecular groups with delocalized orbitals, 
complex structured molecules with three-terminal or multiple-terminal 
configurations and charge-transfer molecular complexes. 
In general, electron-vibronic coupling can be strong in such single-molecule 
devices, whose effects need to be sorted out. The recent surge of activity on 
integrating molecular functionality on a semiconductor platform also brings 
additional functionality through contact 
engineering~\cite{MolFET5,NDR2,Sem1,Sem2,Sem3}  
by attaching the molecule to the surface of bulk semiconductor, semiconductor 
quantum well, quantum wire or quantum dots. 

\section{Discussion and Conclusion}
Central to the vision of nanotechnology is the idea that by developing and following 
a common intellectual path --- the bottom-up paradigm of nanoscale science and 
technology --- it will be possible in the future to assemble virtually any kind of devices 
or functional systems. Much thus lies in the hands of chemists and materials 
scientists, where the goal is to control with atomic precision the morphology, 
structure, composition, and size of the nanoscale building blocks. Next, 
understanding the physics of nanoscale materials emerging from the synthetic 
efforts and inserted into the device and system configurations, i.e., the effect on the 
operating behavior of nanostructures due to the introduction of contact, functional 
interface, the application of external forces and processing/environment- induced 
parameter variations,  is a fundamental part of the bottom-up paradigm, which 
define properties that may ultimately be exploited for nanotechnologies and 
enable us to make rational predictions and define new device concepts unique 
to the nanoscale building blocks.  Finally, to fully exploit the bottom-up paradigm, 
we must develop rational methods of organizing building blocks and device 
elements on multiple length scales. This includes not only assembling building 
blocks in close-packed arrays for interconnectivity but also controlling the 
architecture or the spacing on multiple length scales, i.e., hierarchical assembly, 
which must be done within the context of architectural 
design~\cite{NW1,NWC1,NWC2,NWC3,NWC4,Design1,Design2,Design3}. 

We have focused our attention in this work on materials and memory/logic devices. 
But many of the materials and device structures in molecular electronics 
can be easily configured for applications in chemical/bio- sensors and 
electromechanical devices~\cite{Nano1,Nano2,Ratner,CNT,NW1,NW2}.  
In addition, molecular 
electronics may play an important role in solving the 3-D interconnect problem in 
the ultimately scaled nanoelectronic systems~\cite{Mei1,Mei2,Seeman1,Seeman2}. 
Research progress in molecular electronics systems is steady and strong, which 
gives us cause to believe that funtional molecular electronics systems may be 
practical in ten to fifteen years. Challenges to making this a reality are plentiful 
at every level, some naturally in the fundamental physics and chemistry of 
nanoelectronic materials and devices, but many in architecture and system 
design. These include fabricating and integrating devices, managing their power 
and timing, finding fault-tolerant and defect-tolerant circuits, designing 
and verifying billion-gate systems. Any one of these could block practical 
molecular electronics if unsolved. 

\emph{Acknowledgments}  
We are grateful to the financial support by the DARPA MoleApps program 
and by the ARO/DURINT program. Y.X. has also been supported by the 
MARCO/DARPA Interconnect Focus Center.  M.R. is also supported by the 
NSF Network for Computational Nanotechnology and the NASA URETI program. 
This paper is dedicated to Ned Seeman, visionary and friend. 
%\end{acknowledgments}

%\begin{equation}
%\vec{a}\times\vec{b}=\vec{c}
%\end{equation}

%\subsubsection{Subsubsection Heading}
%Your text goes here. Use the \LaTeX\ automatism for cross-references as
%well as for your citations, see Sect.~\ref{sec:1}.

%\subparagraph{Subparagraph Heading.} Your text goes here.%
%
\index{paragraph}
% Use the \index{} command to code your index words
%
% For tables use
%
%\begin{table}
%\centering
%\caption{Please write your table caption here}
%\label{tab:1}       % Give a unique label
%
% For LaTeX tables use
%
%\begin{tabular}{lll}
%\hline\noalign{\smallskip}
%first & second & third  \\
%\noalign{\smallskip}\hline\noalign{\smallskip}
%number & number & number \\
%number & number & number \\
%\noalign{\smallskip}\hline
%\end{tabular}
%\end{table}
%
%
% For figures use
%
%\begin{figure}
%\centering
% Use the relevant command for your figure-insertion program
% to insert the figure file.
% For example, with the option graphics use
%\includegraphics[height=4cm]{figure.eps}
%
% If not, use
%\picplace{5cm}{2cm} % Give the correct figure height and width in cm
%
%\caption{Please write your figure caption here}
%\label{fig:1}       % Give a unique label
%\end{figure}
%
% For built-in environments use
%
%\begin{theorem}
%Theorem text\footnote{Footnote} goes here.
%\end{theorem}
%
% or
%
%\begin{lemma}
%Lemma text goes here.
%\end{lemma}
%
%
% BibTeX users please use
% \bibliographystyle{}
% \bibliography{}

\begin{thebibliography}{99.}
\bibitem{Note} Although there is no implication on the scaling limit of the 
channel width from the current two-dimensional scaling models, we can 
reasonably expect it to be comparable to the channel length.
\end{thebibliography}
%
% Non-BibTeX users please follow the syntax
% the syntax of "referenc.tex" for your own citations
%\input{xueRef}
\email{yxue@uamail.albany.edu}
\homepage{http://www.albany.edu/~yx152122}
 
%%%%%%%%%%%%%%%%%%%%%%%%%%%%%%%%%%%%%%%%%%%%%%%%%%%%%%%%%%%%%%%%%%%%%%

%%%%%%%%%%%%%%%%%%%%%%%%%%%%%%%%%%%%%%%%%%%%%%%%%%%%%%%%%%%%%%%%%%%%%%

\end{document}